\documentclass[a4paper, 12pt, times]{elsarticle} %

\biboptions{sort&compress}
\usepackage{geometry}
\usepackage{longtable}

\usepackage{amsmath}
\usepackage{siunitx}
\usepackage{lineno,hyperref}
\modulolinenumbers[5]

\makeatletter
\newcommand*{\rom}[1]{\expandafter\@slowromancap\romannumeral #1@}
\makeatother\newcommand*\diff{\mathop{}\!\mathrm{d}}
\usepackage{upgreek}
\usepackage{graphicx}
\usepackage{subfigure}
\usepackage{color}
\usepackage{caption}
\usepackage{tabularx}
\usepackage[T1]{fontenc}

\newcommand \MZ [1]{\bgroup\noindent[\textcolor{blue}{\textbf{MZ}: #1}]\egroup\ignorespacesafterend}

\DeclareMathAlphabet{\Ibb}{U}{msb}{m}{n}

\newcommand{\Be}{{\boldsymbol{\mathnormal e}}}

\newcommand{\Bn}{{\boldsymbol{\mathnormal n}}}
\newcommand{\Bt}{{\boldsymbol{\mathnormal t}}}

\newcommand{\BC}{{\boldsymbol{\mathnormal C}}}

\newcommand{\BM}{{\boldsymbol{\mathnormal M}}}

\newcommand{\BX}{{\boldsymbol{\mathnormal X}}}
\newcommand{\BY}{{\boldsymbol{\mathnormal Y}}}

\newcommand{\Bxi}    {\ensuremath{\boldsymbol\xi}}
\newcommand{\Blambda}   {\ensuremath{\boldsymbol\lambda}}
\newcommand{\Bgamma} {\ensuremath{\boldsymbol\gamma}}
\newcommand{\Bchi} {\ensuremath{\boldsymbol\chi}}

\newcommand{\Bsigma} {\ensuremath{\boldsymbol\sigma}}
\newcommand{\Bepsilon} {\ensuremath{\boldsymbol\epsilon}}

\newcommand{\BGamma} {\ensuremath{\boldsymbol\Gamma}}
\newcommand{\BPhi} {\ensuremath{\boldsymbol\Phi}}
\newcommand{\BPsi} {\ensuremath{\boldsymbol\Psi}}

\newcommand{\Brho}   {\ensuremath{\boldsymbol\rho}}

\renewcommand{\eqref}[1]{Eq.~(\ref{#1})}

\begin{document}

\begin{frontmatter}


\title{Thermodynamic considerations on a class of dislocation-based constitutive models}

\author[NWPU]{Ronghai Wu\corref{corr}}
\cortext[corr]{Corresponding author.}
\ead{ronghai.wu@nwpu.edu.cn}
\author[FAU,SWPU]{Michael Zaiser\corref{corr}}
\ead{michael.zaiser@fau.de}
\address[NWPU]{School of Mechanics, Civil Engineering and Architecture, Northwestern Polytechnical University, Xian, 710072, PR China}
\address[FAU]{WW8-Materials Simulation, Department of Materials Science, Friedrich-Alexander Universit\"at Erlangen-N\"urnberg, Dr.-Mack-Str. 77, 90762 F\"urth, Germany}
\address[SWPU]{Southwest Petroleum University, Department of Materials Science, Chengdu, PR China}

\begin{abstract}
Dislocations are the main carriers of plastic deformation in crystalline materials. Physically based constitutive equations of crystal plasticity typically incorporate dislocation mechanisms, using a dislocation density based description of dislocation microstructure evolution and plastic flow. Typically, such constitutive models are not formulated in a thermodynamic framework. Nevertheless, fundamental considerations 
of thermodynamic consistency impose constraints on the admissible range of model parameters and/or on the range of application of such models. In particular, it is mandatory to ensure that the internal energy increase associated with dislocation accumulation is properly accounted for in the local energy balance. We demonstrate on some examples taken from the literature how failure to do so can lead to constitutive equations that violate the first and second laws of thermodynamics either generally or in some particular limit cases, and we discuss how to formulate constraints that recover thermodynamic consistency.    
\end{abstract}

\begin{keyword}
dislocations; crystal plasticity; constitutive modelling; thermodynamics.
\end{keyword}

\end{frontmatter}

\section*{Nomenclature}
\begin{longtable}{@{}l @{\hspace{5mm}} l }

{\bf General conventions} &\\[12pt]
    $X^s$ \qquad & upper Italic indices enumerate scalar variables $X$\\
    $X_i$ or $X_{ij}$ \qquad & lower Italic indices indicate components of mathematical vectors or tensors\\
    $\BX$ \qquad & bold symbols represent 1) mathematical vectors or tensors (e.g. $\BC$, $\Bsigma$, $\Bepsilon$);\\ &    
2) state vectors that are lists of scalar state variables $X^s$  (e.g. $\Brho$, $\BGamma$, $\Bgamma$),\\&
 lists of scalar multipliers of state variables (e.g. $\Blambda$, $\Bxi$ and $\Bchi$),  or\\&
 matrices describing  couplings between state variables ($\BPsi$, $\BPhi$) \\
    $\BX*\BY = \sum_{s} X^{s}Y^{s}$  \qquad & scalar product of two state vectors $\BX$ and $\BY$ of equal length\\
    $\BX \cdot \BY= \sum_{i} X_{i}Y_{i}$ \qquad & inner product between two mathematical vectors $\BX$ and $\BY$\\
    $\BX:\BY= \sum_{i} \sum_{j} X_{ij}Y_{ij}$ \qquad & double inner product between two mathematical tensors ($\BX$ and $\BY$)\\
    $\dot{X}, \dot{\BX}$ \qquad & temporal evolution rate of $X,\BX$  \\[12pt]

{\bf Specific notations} &\\[12pt]

      $\mu$ \qquad& shear modulus\\
      $b$ \qquad& length of Burgers vector\\
	${\BC}$ \qquad & elastic stiffness tensor \\

	$\Bsigma$ \qquad & stress tensor\\
      $\Bsigma'$ \qquad & deviatoric stress tensor \\
	$\Bsigma_{\rm eq}$ \qquad & equivalent stress \\
      $\sigma_{\rm f}$ \qquad & flow stress \\
	$\sigma_{\rm y}$ \qquad & yield stress\\
      $\tau$ \qquad & resolved shear stress \\
	$\tau_{\rm c}$ \qquad & critical resolved shear stress \\

	${\Bepsilon}$ \qquad & total strain tensor \\
	${\Bepsilon}_{\rm pl} = \sum_s \Gamma^s \BM^s$ \qquad & plastic strain tensor \\
      $\Gamma^s$ \qquad & scalar plastic strain variable (e.g. equivalent strain, shear strain on a slip system) \\
      $\BM^s$ & projection tensor of plastic strain variable  $\Gamma^s$\\
	$\BGamma$ \qquad & state vector compiling plastic strain variables $\Gamma^s$   \\
  	$\gamma^s$ \qquad & resolved (plastic) shear strain onslip system $s$ in crystal plasticity   \\
	$\Bgamma$ \qquad & state vector compiling resolved shear strains $\gamma^s$  in all slip systems \\
      $M$  \qquad & scalar factor (Schmid factor) relating shear strain $\gamma$ and axial strain measure ${\epsilon}_{\rm pl}$  \\

       $\dot{\epsilon}_0$ \qquad & reference strain rate \\
	$\dot{\gamma}_0$ \qquad & reference shear rate \\
      $m$ \qquad & strain rate exponent\\

      $\rho^s$ \qquad & scalar dislocation density, describing (sub)population of dislocation system  \\
	$\Brho$ \qquad & state vector compiling densities  $\rho^s$ of all (sub)populations   \\
      $\Psi^{ss'}(\rho)$  \qquad & dislocation density dependent coupling coefficient between $\dot{\gamma}^{s'}$ and $\dot{\rho}^s$ \\
	$\BPsi(\Brho)$ \qquad & coupling matrix betrween $\dot{\BGamma}$ and $\dot{\Brho}$, elements $\Psi^{ss'}(\rho)$  \\
	$\Phi^s(\rho)$ \qquad & scalar describing static recovery rate of dislocation species $s$\\
      $\BPhi(\Brho)$ \qquad & state vector compiling static recovery rates\\
	$\rho_{\rm S}$ \qquad & density of statistically stored dislocations \\
	$\rho_{\rm G} = \eta/b$ \qquad & density of geometrically necessary dislocations \\
	$\eta$ \qquad & scalar measuring plastic strain gradient \\

      $E_{\rm d}$ \qquad & free energy of dislocation system\\
	$E_{\rm dc}$ \qquad & core energy contribution to dislocation free energy\\
      $e_{\rm dc}^s$ \qquad & core energy per unit line length of dislocation species $s$\\
      $\Be_{\rm dc}$ \qquad & state vector compiling core energy contributions to dislocation line energy \\ 
	$E_{\rm de}$ \qquad & elastic contribution to dislocation free energy\\
      $e_{\rm de}^s = \mu b^2 \lambda_{\rm de}^s$ &  elastic energy per unit line length of dislocation species $s$\\
      $\Be_{\rm de}$ \qquad & state vector compiling elastic energy contributions to dislocation line energy \\ 
      $e_{\rm d}^s = \mu b^2 \lambda^s$& line energy of dislocation species $s$\\
      $\Be_{\rm d} = \mu b^2 \Blambda$ & state vector compiling dislocation line energies\\
      $\chi^{s}$ \qquad & partial derivative of $\Blambda*\Brho$ with respect to $\rho^s$  \\
      $\Bchi$ \qquad & state vector compiling $\chi^s$  \\
	$E_{\rm el}$ \qquad & coarse grained elastic energy\\
	$e_{\rm el} = \Bsigma: (\Bepsilon - \Bepsilon_{\rm pl})$ & coarse grained elastic energy density\\
	$\cal D$ \qquad & local dissipation rate \\

	$L_{\rm dr}$ \qquad & dynamic recovery length \\
	$y_{\rm e}$ \qquad & edge dislocation annihilation distance \\
	$\alpha$  \qquad & coefficient in Taylor friction stress \\
	$\beta$  \qquad & dislocation mean free path in units of dislocation spacings \\
	$d$ \qquad & average grain size \\
	$d_{\rm b}$  \qquad & distance of material point from the nearest grain boundary \\
	$l $  \qquad & storage mean free path \\
	$K$  \qquad & coefficient for dislocation storage at grain boundaries \\

\end{longtable}

\section{Introduction}

Models of plastic deformation fall into two generic classes. On the one hand, phenomenological constitutive equations as used in the solid mechanics community aim at reproducing material behavior observed in experiments by reverse modelling, ie. adjusting the parameters of a given mathematical expression to match computational predictions and experimental observations. Owing to the great freedom of choice in candidate mathematical structures for constitutive equations, the solid mechanics community has always taken great care to restrict the potential range
of constitutive equations by ensuring that they are constructed in a manner that {\em a priori}  satisfies the first and second laws of thermodynamics \cite{ottosen2005mechanics}.

Physically based constitutive models, on the other hand, seek to establish the mathematical structure of constitutive equations with internal variables by associating these variables with physically observable quantities such as dislocation densities. Accordingly, the structure of terms in the constitutive equations is to some extent constrained by the physical processes (e.g dislocation storage, dislocation annihilation) they are supposed to describe. The question of thermodynamic consistency of such models, on the other hand, has received less attention and indeed some models take care {\em not} to restrict the freedom of constitutive modelling by using a formulation which derives constitutive equations from variational principles that involve thermodynamic potentials (an example where this is spelt out explicitly is the conventional gradient plasticity theory of \citet{2004_IJP_Huang}). To be clear, there is nothing wrong with such an approach, only that its thermodynamic consistency must be demonstrated (or contradicted) by independent considerations. In particular, it is self evident that concepts such as a dislocation related contribution to the free energy or entropy of a deforming crystal can be meaningfully formulated and related to the physics of dislocations in all cases where the physical system under consideration is a crystal deforming by dislocation slip. In the present paper, we use such concepts to probe the thermodynamic consistency of some constitutive models.

We focus on a class of models that use continuum descriptions of dislocation microstructure in terms of {\em dislocation densities}. The conceptual problems we are going to discuss are largely absent from discrete dislocation dynamics (DDD) simulations where the implementation of the dynamics is based on configurational forces that derive from the local stress and thereby from the elastic energy functional of the deforming crystal (for a DDD formulation where this is made explicit, see \citet{ghoniem2000parametric}). Several dislocation density based formulations have been based on direct averaging of the DDD dynamics \cite{2003_AM_Groma,2014_JMPS_Geers,valdenaire2016density} and therefore inherit the thermodynamic consistency of the microscopic theory. Others have been derived in a thermodynamically consistent manner from a density-based energy functional which has been either obtained by averaging from the elastic energy functional of the corresponding discrete dislocation system \cite{zaiser2015local} or postulated axiomatically \cite{2015_PRL_Groma, 2016_JMPS_Hochrainer, 2018_PRB_Wu}.

However, there exists a very large number of models which consider the evolution of dislocation densities without explicitly invoking thermodynamic considerations. Starting from the seminal work of \citet{1981_AM_Mecking}, such models use ordinary differential equations to describe the evolution of dislocation densities within the volume of interest. The relationship with plasticity is established, on the one hand, through Orowan-type relations which express the strain rate (in crystal plasticity formulations: the slip rates on the active slip systems) in terms of dislocation densities and velocities. The latter are in turn related to the internal microstructure as described by the dislocation densities and possible additional parameters, and to thermodynamic driving forces for plastic deformation such as equivalent stress or resolved shear stresses. There are literally hundreds of such models in the literature, see eg \cite{2008_IJP_Beyerlein, 2006_AM_Ma,1998_Estrin_JMPT, 2004_AM_Ma, 2010_MM_Luo, 2005_JMPT_Lin, 2010_IJP_Lee, 2011_IJMS_Zhan, 1997_MSEA_Tabourot, 2014_IJP_Knezevic,2004_PM_Arsenlis,2004_JMPS_Evers, 2004_IJSS_Evers,2014_IJP_Li, 2020_JMPS_Zhou, 2008_AM_Kubin, 2020_JMPS_Akhondzadeh, 2000_JMPS_Acharya, 2011_IJP_Fan, 2020_IJP_Zhao, 2020_IJP_Zhang, 2018_IJP_Zheng, 2021_AM_Lieou, 2020_AM_Lim}, so any selection is of necessity biased and uncomplete. Nevertheless these models share common traits and common conceptual constraints which we shall discuss in the following.  

\section{Theoretical framework}

\subsection{Local density-based constitutive models}

In the following we are concerned with models that we may characterize as {\em local dislocation-density based models}. In these models, the evolution of dislocation densities is described by ordinary differential equations. If implemented in a continuum plasticity framework, this implies that dislocation densities are envisaged as local variables that are internal to the volume element under consideration: dislocation transport and the topological nature of dislocations as curved and connected lines are not explicitly considered. We note that such a description is always feasible if only the scale of the elementary volume is large enough such as to exceed typical dislocation transport lengths and loop sizes. 

All such models are built by combination of three different building blocks: 

\begin{enumerate}
\item The dislocation system is characterized in terms of a set of density variables $\rho^{s}$ where the superscript $s$ may, depending on model formulation, distinguish dislocations of different mobility, orientation, and/or slip system. These  densities are understood as describing the line length per unit volume of dislocation lines with the respective characters and can be formally combined into a state vector $\Brho$. In the following upper Italic indices enumerate scalar variables $x^s$ which are formally compiled into state vectors $\BX$. The components of such a state vector $\BX$ transform upon change of coordinate system like scalars. We denote the scalar product of two state vectors $\BX$ and $\BY$ of equal length by an asterisk i.e., $\BX*\BY = x^{s}y^{s} = \sum_{s} x^{s}y^{s}$ where we retain the summation convention for upper indices. Lower Italic indices indicate spatial components of a vector or tensor, eg. $\sigma_{ij}$ and $\epsilon_{ij}$ are the respective components of the stress tensor $\Bsigma$ and total strain tensor $\Bepsilon$ in a Cartesian coordinate system. Here the inner product is marked by $'.'$ and the double inner product by $':'$, e.g., $\Bt_{\Bn} = \Bsigma.\Bn$ is the traction on a surface element of unit normal vector $\Bn$, and $e_{\rm el} = \Bsigma:\Bepsilon_{\rm el}$ is the elastic energy density.
\item Plastic flow is characterized in terms of a set of scalar strain rate variables $\dot{\Gamma}^{\beta}$. Depending on model formulation, the precise meaning of these variables may range from a scalar equivalent strain rate to shear rates on active slip systems. The strain rate variables can be formally combined into a state vector $\dot{\BGamma}$. 
\item The strain rate variables are specified in terms of a set of scalar constitutive equations of the type 
\begin{equation}
\dot{\BGamma} = \dot{\BGamma}(\Brho,\Bsigma)
\end{equation}
which connect the plastic strain rate variables to the local stress state $\Bsigma$ and the local dislocation microstructure. From the strain rate variables the tensorial plastic strain rate can be reconstructed as $\dot{\Bepsilon}^{\rm p} = \sum_{s} \BM^{s}\dot{\Gamma}^{s}$ where the $\BM^{s}$ are direction tensors. 
\item
The evolution of the dislocation densities is specified by a set of {\em kinematic equations} which relate the temporal change of the dislocation densities $\Brho$ to the current dislocation configuration and the plastic strain rate
\begin{equation}
\dot{\Brho} = \BPsi(\Brho)*\dot{\Bgamma} + \BPhi(\Brho)
\end{equation}
where the first, {\em flux driven} terms on the right-hand side, which are proportional to the plastic strain rates, describe irreversible, strain driven dislocation processes such as dislocation multiplication and dynamic recovery. The coupling terms $\Psi^{ss'}$, which can be formally envisaged as elements of the matrix $\BPsi$, describe how the evolution of the dislocation density variables is driven by the strain rate variables: $\dot{\rho}^{s} = \sum_{s'} \Psi^{ss'} \dot{\Gamma}^{s'}$. The flux independent terms $\BPhi$, on the other hand, characterize static recovery processes eg during annealing.
\end{enumerate}

\subsection{Free energy functional of the dislocation system}

We envisage the free energy of a dislocated crystal on a scale well above the single-dislocation spacing, where the dislocation microstructure can be described by dislocation densities. Plastic deformation is associated with two main free energy contributions, namely an elastic energy associated with the applied boundary tractions and the plastic eigenstrain, and a defect free energy associated with the presence of dislocations. The latter encompasses the excess energy of the dislocation cores as well as the elastic energy of the dislocation related eigenstrain fields on scales below the scale of the elementary volume, which is hence to be envisaged as a local elastic energy density associated with the material point. In comparison to core and elastic energy entropic contributions are negligible at all temperatures up to the melting point: Whereas entropic effects may very significantly increase the dislocation nucleation rate under stress \cite{ryu2011entropic}, the contributions of vibrational and configurational entropy to the crystal free energy are negligible at all temperatures up to the melting point \cite{cottrell1953dislocations,forsblom2004vibrational} and accordingly, the thermal equilibrium concentration of dislocations is negligible: Dislocations are no thermal defects. Accordingly, the densities of dislocations are controlled by the mechanical work expended during plastic deformation in their creation, in conjunction with kinematic constraints which prevent their annihilation. 

The free energy of the dislocations contained in a volume V can, on scales well above the dislocation spacing, be expressed as
\begin{equation}
E_{\rm d} \approx E_{\rm dc} + E_{\rm de}
\end{equation}
where, as long as the separation of dislocations is well above their core radius, the dislocation core energy $E^{\rm dc}$ can be expressed in terms of the dislocation density as
\begin{equation}
E_{\rm dc} = \int_V \Brho*\Be_{\rm dc} \diff V
\end{equation}
where the components of $\Brho$ are the densities (dislocation lengths per unit volume) of the dislocation species envisaged in the constitutive formalism, and $\Be_{\rm dc}$ is a vector compiling the respective average dislocation core energies per unit line length. Similarly, the elastic energy of the dislocation stress fields can, at lowest order in the dislocation densities, be expressed as \cite{zaiser2015local}
\begin{equation}
E_{\rm de} = \int_V \Brho*\Be_{\rm de} \diff V
\end{equation}
where $\Be_{\rm de} = \mu b^2 \Blambda_{\rm de}(\Brho)$ The vector $\Blambda_{\rm de}$ compiles nondimensional factors $\lambda_{\rm de}^{s}$ for the dislocation species $\rho^{s}$. These factors are of the order of unity, they depend on the orientation distribution of dislocations and weakly (logarithmically) on the dislocation densities and other geometrical factors characterizing the dislocation arrangements. The elastic energy densities compiled in the 
vector $\Be_{\rm de}$ capture the elastic energy stored in the near-core dislocation stress fields which cannot be resolved in a coarse-grained continuum plasticity theory. To fully describe the energy stored in form of lattice distortions, $E_{\rm de}$ complements the conventional coarse grained elastic energy density 
\begin{equation}
E_{\rm el} = \int_V e_{\rm el} \diff V = \frac{1}{2} \int_V (\Bepsilon-\Bepsilon_{\rm pl}): \BC : (\Bepsilon-\Bepsilon_{\rm pl}) \diff V
\end{equation}
where $\Bepsilon$ and $\Bepsilon_{\rm pl} = \Bepsilon-\Bepsilon_{\rm el}$ are the (coarse grained) total and plastic strain tensors, respectively. Dislocation core and elastic energies can be formally combined such that the free energy functional is divided into a coarse grained part, which depends only on the coarse grained strain fields, and a dislocation related part:
\begin{equation} 
\label{eq: total free energy}
E = \int_V (e_\text{el} + \Be_{\rm d}*\Brho) \diff V,
\end{equation} 
where $\Be_{\rm d} = \mu b^2 \Blambda(\Brho)$ is a vector compiling the dislocation line energies. The corresponding nondimensional factors are compiled into $\Blambda = \Blambda_{\rm de} + \Be_{\rm dc}/(\mu b^2)$; they may depend logarithmically on the dislocation density and/or on geometrical parameters characterizing the dislocation arrangement. 

\subsection{Dissipation inequality with evolving dislocation densities}

The problem of thermodynamic consistency is here formulated in terms of the Clausius-Planck inequality. According to this inequality, the local dissipation rate, which is required to be positively definite, is expressed as 
\begin{equation}
{\cal D} = \Bsigma:\dot{\Bepsilon} - \dot{e}_{\rm el} - \mu b^2 \Bchi*\dot{\Brho} \ge 0
\end{equation}
where $\Bchi$ has the components $\chi^{s} = \lambda^{s} + \sum_{s'}\rho^{s'} \partial \lambda^{s'}/\partial \rho^{s}$. Using that $\Bsigma = \partial e_{\rm e}/\partial \Bepsilon = - \partial e_{\rm el}/\partial \Bepsilon_{\rm pl}$, this is rewritten as 
\begin{equation}
{\cal D} = \Bsigma:\dot{\Bepsilon}_{\rm pl} - \mu b^2 \Bchi*\dot{\Brho} \ge 0
\end{equation}
We now express both the plastic strain rate and the change in dislocation densities in terms of the strain rate variables $\dot{\BGamma}$ to obtain the inequality
\begin{equation} 
\sum_{s} \Bsigma:\BM^{s}\dot{\Gamma}^{s} \ge \mu b^2 \Bchi*(\BPhi(\Brho)*\dot{\BGamma} + \BPsi(\Brho)). 
\end{equation} 
A sufficient condition for this inequality to be fulfilled is that it holds even in the limit when static recovery processes (or more generally diffusion driven processes) are negligible. Thus we consider for the flux driven evolution of the dislocation system the condition
\begin{equation} 
\sum_{s} \Bsigma:\BM^{s}\dot{\Gamma}^{s} \ge \sum_{ss'} \mu b^2 \chi^s \Psi^{ss'}\dot{\Gamma}^{s'}. 
\end{equation}
In general, by change of the loading conditions acting on the local volume element the strain rate variables may vary  in arbitrary proportions. Thus, the inequality must be fulfilled for each term of the sum separately:
\begin{equation} 
\dot{\Gamma}^{s} [ \Bsigma:\BM^{s}  - \mu b^2 \sum^{s'} \chi^{s'} \Psi^{ss'}] \ge 0
\label{eq:dissipation_ineq}
\end{equation}
This inequality can be fulfilled in two manners: either the strain rate variables $\dot{\Gamma}^{s}$ are zero, or the corresponding terms in the brackets must be positively definite. (A further alternative would be to consider the line energy factors $\chi^{s}$ to be negligible, which we deem unphysical). 

In the following we first use Eq \ref{eq:dissipation_ineq} to probe the thermodynamic implications of defect accumulation during dislocation hardening for some models published in the literature. We then propose a generic method to formally ensure thermodynamic consistency for local dislocation based plasticity models. 

\section{Examples of local dislocation density based models}

\subsection{The model of Kocks and Mecking}

The seminal paper of \citet{kocks1976laws,1981_AM_Mecking} can be considered at the origin of most currently used dislocation density based hardening models. Here we refer to the original version of \citet{kocks1976laws}. The model is conceived for the analysis of uni-axial tensile tests with axial stress $\sigma$. It considers a single scalar dislocation density variable $\rho$ and a single strain rate variable $\dot{\gamma}$. The function $\Psi(\rho)$ is taken in the form
\begin{equation} 
\Psi(\rho) = {\rm sign}(\dot{\gamma})\tilde{\Psi}(\rho), \quad \tilde{\Psi}(\rho)=\frac{1}{b\beta} \sqrt{\rho} - \frac{L_{\rm dr}}{b} \rho
\label{eq:KM}
\end{equation}
where $L_{\rm dr}$ is a dynamic recovery length (dislocation annihilation cross-section) and the factor $\beta$ gives the dislocation mean free path in units of dislocation spacings. In Eq. (\ref{eq:KM} we have introduced the sign function 
\begin{equation}
{\rm sign}(x) = \left\{
\begin{array}{ll}
1&,\quad x>0,\\
0&,\quad x=0,\\
-1&,\quad x<0.
\end{array}
\right.
\end{equation}
which is not explicitly stated in the original papers, which always envisage monotonic deformation in tension. This function ensures that the evolution of dislocation density, which according to the model first increases and then saturates during deformation, does not depend upon the direction of straining and is not reversed upon stress reversal: Kocks-Mecking is a model for isotropic hardening. 

The relationship between the strain rate variable, axial stress and dislocation density is assumed in the form
\begin{equation}
\dot{\gamma} = {\rm sign}(\sigma)\dot{\gamma}_0 \left(\frac{M|\sigma|}{\alpha \mu b \sqrt{\rho}}\right)^m
\end{equation}
Here $M$ is an average orientation factor such that the axial plastic strain rate is given by $\dot{\epsilon}_{\rm pl} = M\dot{\gamma}$. Using these expressions, the dissipation inequality can be written as 
\begin{equation} 
\dot{\Gamma}[ M\sigma - \mu b^2 \chi \Psi] \ge 0
\end{equation}
It is easy to see that for small $\rho$, when the dynamic recovery term in Eq (\ref{eq:KM}) can be neglected, this inequality is violated once the strain rate magnitude falls into the interval $0 < |\dot{\gamma}| < \dot{\gamma}_0 [\chi/(\alpha\beta)]^m$.

To be fair, we should note that the aim of Kocks and Mecking has never been to formulate a thermodynamically consistent continuum plasticity model but to develop a framework for analysing hardening curves: Their work is posited in the domain of physical metallurgy rather than continuum mechanics of materials. Indeed, for typical values of the parameters that match empirical hardening data in fcc crystals, the ratio $\chi/(\alpha\beta)$ is a small number whereas the exponent $m$ is large. Accordingly, the strain rates in the offending strain rate regime are exceedingly small and, hence, the problem is of conceptual rather than numerical nature.  However, the Kocks-Mecking model has served as a template for constructing many other constitutive models, some of which suffer from similar problems to a larger degree, as now illustrated.

\subsection{The model of Lefebvre et al.}

Based on 2D DDD simulations, \citet{lefebvre2007yield} developed a model of dislocation accumulation and Hall-Petch behavior in ultrafine-grained materials. The simulations are not of our concern here, rather we focus on the analytical model. The model uses a scalar plastic strain variable $\epsilon_{\rm p}$ (again, we may think of the axial strain in a uniaxial test) and a scalar dislocation density $\rho$. The  rate independent flow stress is assumed to fulfill the relation
\begin{equation}
\sigma = \sigma_0 + \frac{1}{M} \alpha \mu b \sqrt{\rho}
\end{equation}
and the dislocation storage rate is related to the grain size $d$ via 
\begin{equation}
\partial_t \rho = \frac{5}{db} \dot{\epsilon}_{\rm pl},
\end{equation}
hence, $\Psi(\rho) = 5/(db)$ which, in physical terms, means that the dislocation mean free path to storage is proportional to the grain size. The plastic strain rate is, in the spirit of a rate independent plasticity model, not specified, however, it is tacitly assumed that both the (axial) stress $\sigma$ and the strain rate $\dot{\epsilon}_{\rm pl}$ are positive. From this model the authors conclude that, since $\rho = 5 \epsilon_{\rm pl}/(db)$, a Hall-Petch type proportionality between flow stress and inverse square root of grain size. 

Inserting into the dissipation inequality gives
\begin{equation} 
\dot{\epsilon}_{\rm pl} [ \sigma_0 + \frac{1}{M} \alpha \mu b \sqrt{\rho} -  \frac{5\chi \mu b}{d} ] \ge 0
\label{eq:reduced_dissipation_ineq}
\end{equation}
Irrespective of the magnitude of $\sigma_0$ this inequality is, for positive stress and strain rate, fulfilled as long as $\epsilon_{\rm pl} \ge 5 M^2\chi^2 b/(d\alpha^2)$. Assuming a typical plastic strain offset of $0.2\%$, $\alpha = 0.25$, Burgers vector $b = 0.25$nm,  $M =1/3$, and $\chi \approx 1$ this implies that the model works as long as the grain size is above 1.1 $\mu$m. However, at smaller strains ($\epsilon_{\rm pl} \to 0$) there is always an inconsistency unless one assumes that $\sigma_0 \ge 5\chi \mu b/d$. Hence, the leading order term of the Hall-Petch-like relationship for small grain sizes must, if one otherwise follows the argument of \citet{lefebvre2007yield}, be proportional to $d^{-1}$ rather than $d^{-1/2}$. 

\subsection{The model of Haouala et al.}

The model proposed by \citet{haouala2018analysis,haouala2020effect} can be considered a development of the previous model. Whereas \citet{lefebvre2007yield} consider a coarse grained scale where the elementary volume averages over many grains, \citet{haouala2018analysis,haouala2020effect} use a crystal plasticity framework where plastic deformation is resolved below the grain scale. The model considers slip system specific dislocation densities $\rho^{s}$, hence $s$ is in that model a slip system index. Accordingly, the strain rate variables are shear strain rates on the different slip systems which are assumed of the form
\begin{equation}
\dot{\gamma}^{s} = \dot{\gamma}_0 \left(\frac{|\tau^{s}|}{\tau_{\rm c}^{s}}\right)^m {\rm sign}(\tau^{s})
\end{equation}
where $\tau^{s} = \Bsigma:\BM^{s}$ is the resolved shear stress and the $\BM^s$ are slip system projection tensors. 
The stresses $\tau_{\rm c}^{s}$ are assumed of the form
\begin{equation}
\tau_{\rm c}^{s} = \mu b \sqrt{\sum_{s'} q^{ss'} \rho^{s'}}.
\end{equation}
where the $q^{ss'}$ are coupling constants that define the latent hardening matrix. 
The overall plastic strain rate is in the here used small strain formulation given by
\begin{equation}
\dot{\Bepsilon}_{\rm pl} = \sum_{s}\BM^{s} \dot{\gamma}^{s}
\end{equation}
We note that \citet{haouala2020effect} use a finite-strain framework, which however does not change the conceptual problems which we here formulate in the small strain limit. Dislocation storage is treated in generalization of the models of Kocks-Mecking and Lefebvre et. al. by setting 
\begin{equation}
\partial_t \rho^{s} = \Psi^{s}\dot{\gamma}^{s} \quad,\quad
\Psi^{s}= \frac{{\rm sign}(\dot{\gamma}^{s})}{b} \left(\frac{1}{l^s} - \rho^{s}L_{\rm dr}
\right)
\end{equation}
where the dynamic recovery cross-section is taken to be twice the edge dislocation annihilation distance, $L_{\rm dr} = 2 y_{\rm e}$, and the storage mean free path on slip system $s$ is evaluated in generalization of both Kocks and Lefebvre et al. as 
\begin{equation}
\frac{1}{l^{s}} = {\rm max}\left(\frac{1}{\beta}\sqrt {\sum_{s'} \rho^{s'}}, \frac{K}{d_{\rm b}}\right)
\end{equation}
where $K=5$ in line with Lefebvre et al. $d_{\rm b}$ is understood as the distance between the material point and the nearest grain boundary. 

For single crystals ($d_{\rm b} \to \infty$) the model reduces to a multi-slip generalization of the Kocks-Mecking framework and the same comments apply. However, in the case of finite crystallite size the problem of thermodynamic consistency is now greatly exacerbated. This can be seen by inserting into the dissipation inequality which in the limit of low dislocation densities (no appreciable dynamic recovery) now reads
\begin{equation} 
\dot{\gamma}^{s}(\tau^{s} -  \mu \frac{b}{l^{s}}) \ge 0.
\end{equation} 
Since $l^{s} \to 0$ as $d_{\rm b} \to 0$, for material points sufficiently close to a grain boundary the dissipation inequality is thus violated whatever the deformation parameters. 

\subsection{Conventional mechanism based strain gradient plasticity} 

Conventional mechanism based strain gradient plasticity (CMSG) as popularized by \citet{gao1999mechanism,huang2004conventional} provides a simple (and very popular) generalization of dislocation-based plasticity theory that accounts for strain gradient effects and, hence, for size-dependent plastic deformation behavior. 

The model considers a scalar plastic strain rate variable $\dot{\epsilon}_{\rm pl}$ which relates, in the spirit of classical von Mises plasticity,  to the tensorial plastic strain rate via 
\begin{equation}
\dot{\Bepsilon}_{\rm pl} = \BM(\Bsigma) \dot{\epsilon}_{\rm pl} = \frac{3 \Bsigma'}{2 \sigma_{\rm eq}}\dot{\epsilon}_{\rm pl}
\end{equation}
where $\Bsigma'$ is the deviatoric stress and $\sigma_{\rm eq} = \sqrt{(3/2)\Bsigma':\Bsigma'}$ is the equivalent stress. The strain rate variable is assumed of the form
\begin{equation}
\dot{\epsilon}_{\rm pl} = \dot{\epsilon}_0 \left(\frac{\sigma_{\rm eq}}{\sigma_{\rm f}}\right)^m
\end{equation}
where rate independent plasticity is recovered by making the exponent $m$ very large. The flow stress $\sigma_{\rm f}$ is related to two dislocation density variables, namely a density $\rho_{\rm S}$ of 'statistically stored' dislocations and a density $\rho_{\rm G}$ of 'geometrically necessary' dislocations:
\begin{equation}
\sigma_{\rm f} = \frac{1}{M}  \alpha \mu b \sqrt{\rho_{\rm G} + \rho_{\rm S}}.
\label{eq:taylor}
\end{equation}
where $M \approx 1/3$ for isotropic polycrystals. The 'geometrically necessary' dislocation density is related to a scalar measure $\eta$ of the plastic strain gradient by $\rho_{\rm G} = 
\eta/b$ and accordingly its rate of evolution is
\begin{equation}
\dot{\rho}_{\rm G} = \frac{\dot{\eta}}{b}
\end{equation}
The evolution of the 'statistically stored' density is obtained phenomenologically from observations in a test without strain gradients (e.g. a tensile test on a long thin specimen). Using a positively definite equivalent plastic strain measure $\epsilon_{pl}$, the observed flow stress evolution is $\sigma_{\rm f} = \sigma_{\rm y} f(\epsilon_{\rm pl})$ and the corresponding plastic strain derivative $h_{\rm pl} = \sigma_{\rm y} f'(\epsilon_{\rm pl}$ where $f'= \partial f/\partial \epsilon_{\rm pl}$. With these notations we find from Eq. (\ref{eq:taylor})
\begin{equation}
\dot{\rho}_{\rm S} = \dot{\epsilon}_{\rm pl} \phi(\epsilon_{\rm pl})\quad,\quad \phi(\epsilon_{\rm pl}) = \frac{2M^2 \sigma_y ff'}{(\alpha\mu b)^2}.
\end{equation}
We can now formulate the dissipation inequality, Eq (\ref{eq:dissipation_ineq}), as
\begin{equation} 
\dot{\epsilon}_{\rm pl}\left(1- \frac{2M^2 \chi_{\rm s} \sigma_{\rm y}^2 ff'}{\sigma_{\rm eq}\alpha^2\mu} 
- \frac{\mu b \chi_{\rm g}}{\sigma_{\rm eq}} \frac{\dot{\eta}}{\dot{\epsilon}^{\rm pl}}\right) \ge 0
\end{equation}
where $\chi_{\rm s}$ and $\chi_{\rm g}$ are the line energy coefficients for statistically stored and geometrically necessary dislocations, respectively. 

We now note that the second term in the brackets is of the order of $\sigma_{\rm y}/\mu \ll 1$ such that, in the absence of strain gradients, the dissipation inequality is always fulfilled. This is to be expected, since we simply recover standard phenomenological von Mises plasticity. The thermodynamic consistency requirement thus reduces to 
\begin{equation} 
\frac{\mu b \chi_{\rm g}}{\sigma_{\rm eq}} \frac{\dot{\eta}}{\dot{\epsilon}_{\rm pl}} < 1
\end{equation}
To analyze this expression, we consider the rate-independent limit $m\to\infty$ where $\sigma_{\rm eq} = \alpha \mu b \sqrt{\rho_{\rm S}+\rho_{\rm G}}/M$. We further note that the ratio of the equivalent strain gradient rate $\dot{\eta}$ and the equivalent strain rate $\dot{\epsilon}_{\rm p}$ defines a length $l_{\rm pl} = \dot{\epsilon}_{\rm pl}/\dot{\eta}$ which can be envisaged as the characteristic length of emergent variations of the plastic strain field. The dissipation inequality then states that this length must fulfill the inequality
\begin{equation} 
l_{\rm pl} > \frac{M\chi_{\rm G}}{\alpha \sqrt{\rho_{\rm S}+\rho_{\rm G}}}.
\end{equation}
Since $M\chi_{\rm G}/\alpha$ is of order of one, this inequality is tantamount to the requirement that the characteristic length scale of plastic strain variations must be larger than the average dislocation spacing. This requirement, which here comes in through the thermodynamic back door, is physically meaningful: it is self evident that a continuum description of dislocation plasticity cannot be used to describe
structures of the plastic strain field below the spacing of the discrete dislocations. We can thus conclude that CMSG is always thermodynamically consistent as long as it is used within its intrinsic limits of applicability. On the other hand, applying CMSG to systems which are so small as to initially contain few or no dislocations not only overstretches the conceptual foundations of the theory but is also thermodynamically not acceptable.

\section{Thermodynamically consistent dislocation density based models}

To construct models that ensure thermodynamic consistency while retaining the framework of a given local dislocation density based model, we observe that the lhs of the dissipation inequality for the strain rate variable $\dot{\Gamma}^s$ can be formally considered as the product of this variable and an associated effective force $T^s$: 
\begin{equation} 
\dot{\Gamma}^{s} \left(\Bsigma:\BM^{s} - \mu b^2 \sum^{s'} \chi^{s'} \Psi^{ss'}\right) 
=: \dot{\Gamma}^{s} T^{\rm s} \ge 0. 
\end{equation}
To proceed further we note that $\Psi^{ss'} = \Psi_{\rm rev}^{ss'} +  {\rm sign}\dot{\gamma}^s\Psi_{\rm irr}^{ss'} $ may consist of contributions describing reversible and irreversible dislocation storage processes, respectively. 

Reversible dislocation storage means that the associated dislocation line energy can be recovered upon reverse straining, e.g. in the case of dislocations emitted from a source and forming a dislocation pile up against a grain boundary: upon strain reversal, these dislocations may return to the source. Accordingly, for a given dislocation state, the terms $\dot{\Gamma}^{s} \Psi_{\rm rev}^{ss'}$ change their sign upon change in sign of the strain rate variable. The associated stress $T_{\rm b}^s = \mu b^2 \sum^{s'} \chi^{s'} \Psi_{\rm rev}^{ss'}$ has the character of a back stress and leads to kinematic hardening effects.

Irreversible dislocation storage processes occur at a rate that does not depend on the sign of straining. Examples are the deposition of activated dislocations in immobile configurations that do not dissolve when the direction of straining is reversed. The associated stress $T_{\rm f}^s = \mu b^2 \sum^{s'} \chi^{s'} \Psi_{\rm irr}^{ss'} {\rm sign}\dot{\gamma^s}$ always opposes the strain rate: it has the character of a friction stress. 

The effects of friction and back stresses can be conveniently incorporated into a generic framework by defining effective driving stresses
\begin{equation}
T_{\rm eff}^s = \left\{
\begin{array}{ll}
\Bsigma:\BM^s - T_{\rm b}^s - T_{\rm f}^s,&\; \Bsigma:\BM^s - T_{\rm b}^s > 0,\\ 
                                          &\; |\Bsigma:\BM^s - T_{\rm b}^s| > T_{\rm f}^s\\
\Bsigma:\BM^s - T_{\rm b}^s + T_{\rm f}^s,&\; \Bsigma:\BM^s - T_{\rm b}^s < 0, \\
                                          &\; |\Bsigma:\BM^s - T_{\rm b}^s| > T_{\rm f}^s\\
0,&\; |\Bsigma:\BM^s - T_{\rm b}^s| \le T_{\rm f}^s
\end{array}
\right.
\end{equation}
Thermodynamic consistency can then be ensured by simply defining the strain rate variables as sign-preserving functions of the associated effective driving stresses.

\section{Discussion and Conclusions}

The main message of this short investigation can be summarized as follows: If one formulates models that describe 
strain hardening in terms of dislocation density accumulation, it is mandatory to properly account for the fact that this process may implies a plastic strain induced increase of the internal energy of the crystal. As a consequence, one has to make sure that the expended work is sufficient to account for this energy increase (hence, the yield stress must be sufficiently high). If one fails to do so, energy conservation can only be maintained by negative dissipation, thus violating the second law. 

This simple truism is not always taken into account by authors who formulate dislocation density based plasticity models. Our selection of examples is both subjective and incomplete, and we should note that many other authors have made similar oversights. As an example we may take a recent paper by one of the present authors \cite{zhao2021tension} where, in the framework of a much more complex model than those analyzed here, a grain size dependent dislocation storage term $~ d^{-1}$ is combined with a Hall-Petch type grain size dependence of the flow stress, $\sigma_{\rm f} ~ d^{-1/2}$. It is easy to see that, for small grain sizes, this model might run into exactly the same problem as described above in Section 3.2. The deeper message here is that, in formulation of dislocation density based constitutive models, the flow stress and the kinematics of dislocation density accumulation cannot be modelled independently but are connected by a thermodynamic consistency requirement. 

A generic concern arises regarding the widespread use of phenomenological power law expressions for strain rate variables, akin to $\dot{\epsilon}_{\rm pl} = \dot{\epsilon}_0 (\sigma/\sigma_{\rm f})^m$ or variations thereof. In the context of deformation by dislocation glide such expressions may turn out to be problematic because they allows, in principle, plastic flow to occur at arbitrarily low stresses and, hence, arbitrarily low work expenditure ({\em 'panta rhei'}). If such flow entails dislocation density accumulation, this is physically impossible.

The present arguments have been formulated for dislocation density accumulation that is kinematically irreversible: the rates of accumulation are assumed to be governed by the moduli of the plastic strain rate variables, hence accumulation does not revert upon reverse straining. Thus the present considerations apply to dislocation-mediated isotropic hardening. Of course, parts of the dislocation density evolution may be kinematically reversible (inversive) \cite{zhang2019microplasticity} and revert upon reverse loading. In this case, the same will be true for the stored defect energy. Defect energy recovery may allow plastic flow to occur at reduced stresses, leading to kinematic hardening effects which 
warrant a separate investigation. 

\section*{Acknowledgment}
The research is supported by Opening Project of Applied Mechanics and Structure Safety Key Laboratory of Sichuan Province (SZDKF-1803), National Natural Science Foundation of China (12002275) and Natural Science Foundation of Shaanxi Province (2020JQ-125).

\bibliography{references}

\end{document}